\documentclass[twoside,12pt]{article}
\usepackage[dvips]{graphics,epsfig}
\usepackage[latin1]{inputenc}
\usepackage{setspace}
\usepackage{textcomp}
\usepackage{amsmath}
\usepackage{graphicx}
\usepackage{epstopdf}
\usepackage{subfig}

\def\laq{\raise 0.4 ex \hbox{$<$}\kern -0.8 em\lower 0.62 ex\hbox{$\sim$}}
\def\gaq{\raise 0.4 ex \hbox{$>$}\kern -0.7 em\lower 0.62 ex\hbox{$\sim$}}

\def\beq{\begin{equation}}
\def\eeq{\end{equation}}
\def\beqa{\begin{eqnarray}} 
\def\eeqa{\end{eqnarray}}

\begin{document}
\pagestyle{plain}

\begin{flushright}
May 31, 2011
\end{flushright}
\vspace{15mm}

\begin{center}

{\Large\bf Ho\v{r}ava-Lifshitz Quantum Cosmology}

\vspace*{1.0cm}

Orfeu Bertolami$^{*}$\\
\vspace*{0.5cm}
{Departamento de F\'\i sica e Astronomia \\ Faculdade de Ci\^{e}ncias da
Universidade do Porto, \\
Rua do Campo Alegre 687, 4169-007 Porto, Portugal. }\\

\vspace*{1.0cm}

Carlos A. D. Zarro$^{**}$\\
\vspace*{0.5cm}
{Centro de F\'\i sica do Porto \\ Departamento de F\'\i sica e Astronomia \\ Faculdade de Ci\^{e}ncias da
Universidade do Porto, \\
Rua do Campo Alegre 687, 4169-007 Porto, Portugal. }\\

\vspace*{2.0cm}
\end{center}

\begin{abstract}
In this work, a minisuperspace model for the projectable Ho\v{r}ava-Lifshitz (HL) gravity without the detailed balance condition is investigated. The Wheeler-deWitt equation is derived and its solutions are studied and discussed for some particular cases where, due to HL gravity, there is a ``potential barrier'' nearby $a=0$. For a vanishing cosmological constant, it is found a normalizable wave function of the universe. When the cosmological constant is non-vanishing, the WKB method is used to obtain solutions for the wave function of the universe. Using the Hamilton-Jacobi equation, one discusses how the transition from quantum to classical regime occurs and, for the case of a  positive cosmological constant, the scale factor is shown to grow exponentially, hence recovering the GR behaviour for the late universe.

\noindent

\end{abstract}

\vfill
\noindent\underline{\hskip 140pt}\\[4pt]
{$^{*}$ Also at Instituto de Plasmas e Fus\~{a}o Nuclear, Instituto Superior T\'ecnico, Av. Rovisco Pais 1, 1049-001
Lisbon,Portugal} \\
{E-mail address: orfeu.bertolami@fc.up.pt} \\
\noindent
{$^{**}$ E-mail address: carlos.zarro@fc.up.pt}

\newpage


\doublespacing

\section{Introduction} \label{sec:Introduction}

Ho\v{r}ava-Lifshitz (HL) gravity is a quite original proposal for a ultraviolet (UV) completion of General Relativity (GR) \cite{Horava:2009uw}, in which gravity turns out to be power-countable renormalizable at the UV fixed point. GR is supposed to be recovered at the infra-red (IR) fixed point, as the theory goes from high-energy scales to low-energy scales. In order to obtain a renormalizable gravity theory one abandons Lorentz symmetry at high-energies \cite{Horava:2009uw,Visser:2009fg}. Even though the idea that the Lorentz symmetry is a low-energy symmetry has been previously considered \cite{Chadha:1982qq}, the novelty of the HL proposal is that the breaking of Lorentz symmetry occurs the very way as in some condensed matter models (cf. Ref. \cite{Horava:2009uw} and references therein), that is through an anisotropic scaling between space and time, namely $\vec{r}\rightarrow b\vec{r}$ and $t \rightarrow b^{z} t$, $b$ being a scale parameter. The dynamical critical exponent $z$ is chosen in order to ensure that the gravitational coupling constant is dimensionless, which makes possible a renormalizable interaction. As the Lorentz symmetry is recovered at the IR fixed point, $z$ flows to $z=1$ in this limit.

The anisotropy between space and time leads rather naturally to the well known $3+1$ Arnowitt-Deser-Misner (ADM) splitting \cite{Arnowitt:1962hi}, originally devised to express GR in a Hamiltonian formulation. Following Ref. \cite{Horava:2009uw}, a foliation,  parametrized by a global time $t$, is introduced. Since the global diffeomorphism is not valid anymore, one imposes a weaker form of this symmetry, the so-called  {\it foliation-preserving diffeomorphism}. Choosing this approach, the lapse ADM function, $N$, is constrained to be function only of the time coordinate, $i.e.$ $N=N(t)$. This assumption satisfies the {\it projectability condition} \cite{Horava:2009uw}. In order to match GR, one could also choose $N=N(\vec{r},t)$, a model dubbed {\it non-projectable} and which has been investigated in Refs. \cite{Blas:2009qj,Blas:2010hb}. The next step involves getting a gravitational Lagrangian into this anisotropic scenario. For this purpose, the  effective field theory (EFT) formalism is used: every term that is marginal or relevant at the UV fixed point ($z\neq 1$) is included and,  at the IR fixed point, only $z=1$ terms survives. GR is then presumably recovered. The number of terms that must be included splits HL gravity into two classes, depending on whether one adopts the {\it detailed balance condition} or not. It is argued in Ref. \cite{Horava:2009uw}, that if one allows every relevant term to be included into the Lagrangian, the number of coupling constants would be so large that any analysis would become impracticable. The detailed-balance condition is inspired by non-equilibrium thermodynamics \cite{Horava:2011gd} and, loosely speaking, it states that the potential terms of a $D$-dimensional action are obtained using a $(D-1)$-dimensional function, the superpotential. It is argued that although detailed balance is a simplifying assumption, it is by no means a necessary one \cite{Sotiriou:2009gy,Sotiriou:2009bx}. It is shown that the list of allowed  terms is not so large after all and the detailed balance Lagrangian is obtained after a proper choice of coefficients. 

A common problem plaguing all HL versions is the presence of a scalar field mode, which has a trans-Bogoliubov dispersion relation with $\vec{k}^{6}$ term \cite{Sotiriou:2010wn,Visser:2011mf}. This scalar mode appears, likewise a Goldstone boson,  after the breaking a continuous symmetry. To avoid this mode, one has to introduce more symmetries: besides the foliation-preserving diffeomorphism, a local $U(1)$ symmetry can be introduced \cite{Horava:2010zj} and it is shown  that the scalar mode is then gauged away. This version of HL gravity is referred to as {\it general covariant}, given that the number of degrees of freedom matches the one of GR.  For reviews on these versions of HL gravity, the reader is referred to Refs. \cite{Horava:2011gd,Sotiriou:2010wn,Visser:2011mf}.      

Cosmological considerations have been extensively studied in the context of HL gravity (for reviews, see Refs. \cite{Mukohyama:2010xz,Saridakis:2011pk}). One subtle point that arises in the projectable version concerns the lapse function which being just a function of time, implies that the classical Hamiltonian constraint of GR is no longer local, and must be integrated over spatial coordinates. It is shown in Ref. \cite{Mukohyama:2009mz} that this yields an additional term that mimics dust into the Friedman equations. However, as noted in Ref. \cite{Sotiriou:2009bx}, the Robertson-Walker metric is homogeneous, so this spatial integral is simply the spatial volume of the space and hence this ``dark dust'' constant must vanish \cite{Maeda:2010ke}. The presence of terms higher spatial curvature terms in HL gravity, gives rise to a plethora of new  cosmologies, that exhibit, for instance, some bouncing and oscillating solutions \cite{Kiritsis:2009sh,Brandenberger:2009yt,Calcagni:2009ar}. A classification of Friedman-Lema\^{i}tre-Robertson-Walker (FLRW) cosmologies for HL gravity has been performed in Refs. \cite{Maeda:2010ke,Wang:2009rw}. One should notice that the analysis carried out in these references is entirely classical and based on the resulting Friedman-like and Rauchaudhury-like equations.    

Quantum cosmology is an interesting step towards the understanding of quantum gravity and the initial conditions of the universe. Its setup consists  in splitting space-time, using the ADM formalism and applying well known quantum mechanical considerations for constrained systems. The cosmological principle is evoked so that space-time is foliated in leaves with a constant global time. To implement the quantum scenario, one promotes the Hamiltonian constraint $H=0$ to an operator equation, the Wheeler-deWitt (WdW) equation, $\hat{H}\psi=0$, where $\psi$ is the wave function of the universe \cite{DeWitt:1967yk}.    The WdW equation is an hyperbolic equation on the space of all 3-metrics, the so-called superspace. Its complexity makes the task of obtaining solutions a formidable one. To deal with this equation, one often considers simpler spaces, such as for instance the FRLW space-times, which leads to a minisuperspace model, where the number of degrees of freedom is considerably reduced from infinite (any 3-metrics) to one (the scale factor) \cite{Hartle:1983ai}. Despite their relative simplicity the minisuperspace models are not completely free from problems. Indeed, one can point out, for instance, the fact that the wave function of the universe is not in many cases normalizable, which implies that the usual interpretation of quantum mechanics cannot be used. However, in the context of some particular models normalizable wave functions have been found and discussed \cite{Kiefer:2010zz,Kamenshchik:2007zj,Bastos:2009kp}. For, comprehensive reviews, see $e.g.$ Refs. \cite{Halliwell:2009rb,Kiefer:2007}.

We argue that quantum cosmology allows for a valuable insight of HL gravity in the quantum context. In both approaches, one foliates the space-time in constant global time leaves, a procedure that automatically satisfy the projectability condition. But when adopting the QC formalism in HL gravity, one faces the problem of turning the Hamiltonian constraint into the WdW equation, since the Hamiltonian constraint in the HL gravity is not local. Nevertheless, choosing a FLRW metric minisuperspace model or, more generally, a spatially homogeneous cosmological metric,  one can argue that the spatial integration  yields a local constraint. Notice that another suitable feature of HL gravity, is that it does not introduce higher than first order time derivatives of the scale factor on the action, making the quantization procedure straightforward as a mixture between time and spatial derivatives and powers of momentum is not found. Indeed, the kinetic part of the Hamiltonian has the same structure as the one in the usual QC, and HL gravity introduces only  higher spatial derivatives terms, which dominate on the very small scales. Notice that the problem of high order time derivatives imposes severe obstacles for applying QC on high order derivatives gravities and string theory, but not for HL gravity.  One concludes then that the minisuperspace model can be naturally implemented in the HL gravity proposal.  In the minisuperspace model, the WdW equation for the HL gravity was obtained in Ref. \cite{Garattini:2009ns}, however in there the interest  was on the cosmological constant problem and the HL WdW equation was neither discussed nor solutions were presented.

In this work, one investigates the projectable HL gravity without detailed balance in the context of the minisuperspace model of quantum cosmology for a FLRW universe without matter. This particular choice, despite being much simpler than the non-projectable version \cite{Blas:2009qj,Blas:2010hb} and the general covariant approach \cite{Horava:2010zj}, exhibits the main features of the HL gravity and contains the detailed balance as a limiting case. A matter sector is not introduced, given that the main interest in the very early universe, where the HL new terms dominate and for the late universe, an epoch dominated by the cosmological constant. Moreover, the inclusion of the matter sector and how it is coupled to HL gravity remains an open question \cite{Sotiriou:2010wn}.

This paper is organized as follows: in section \ref{sec:WdWEquation}, the minisuperspace HL is presented and the WdW-equation is obtained. In section \ref{sec:WdWEquationSolutions}, the solutions of the WdW are found and discussed. In section \ref{sec:interpretation}, the wave function of a HL is interpreted and an analysis of the Hamilton-Jacobi equation is performed. Finally, concluding remarks are presented in section \ref{sec:conclusions}.  


\section{The Wheeler-DeWitt equation}\label{sec:WdWEquation}

\subsection{Metric}
One considers the RW (Robertson-Walker) metric with
$R\times S^{3}$ topology

\begin{equation}\label{eq:FLRWmetric}
 ds^{2}=\sigma^{2}\left(-N(t)^{2}dt^{2}+a^{2}\gamma_{ij}dx^{i}dx^{j}\right)
\end{equation}

\noindent where $i,j=1,2,3$, $\sigma^{2}$ is a normalization constant, $N(t)$ is the lapse function and $\gamma_{ij}$ is the metric of the unit $3$-sphere. Its metric is given by
 $\gamma_{ij}=\mbox{diag}\left(\frac{1}{1-r^ {2}}, r^{2}, r^{2}\sin^{2}\theta \right)$.

The extrinsic curvature takes the form:

\begin{equation}\label{eq:extrinsiccurvature}
 K_{ij}=\frac{1}{2\sigma N}\left(-\frac{\partial g_{ij}}{\partial t}+\nabla_{i}N_{j}+\nabla_{j}N_{i}\right),
\end{equation}

\noindent where $N^{i}$ is the ADM shift vector and $\nabla_{i}$ denoes the $3$-dimensional covariant derivative. As $N_{i}=0$ for RW-like spaces in study,
 
\begin{equation} \label{eq:Kij}
 K_{ij}=-\frac{1}{\sigma N} \frac{\dot{a}}{a} g_{ij}.
\end{equation}

\noindent Taking the trace one gets

\begin{equation} \label{eq:K}
 K=K^{ij}g_{ij}=-\frac{3}{\sigma N} \frac{\dot{a}}{a}.
\end{equation}

\noindent The Ricci components of the $3$-metric can also be obtained, as the foliation is a surface of maximum symmetry

\begin{equation} \label{eq:3metric}
R_{ij}=\frac{2}{\sigma^{2}a^{2}}g_{ij},
\end{equation}

\begin{equation} \label{eq:3metrica}
R=\frac{6}{\sigma^{2}a^{2}}. 
\end{equation}

\subsection{Ho\v{r}ava-Lifshitz action}

The action for the projectable HL gravity without detailed balance  is given by \cite{Sotiriou:2009gy,Sotiriou:2009bx}:

\begin{equation}\label{eq:HLaction1}\begin{split}
S_{HL}&=\frac{M_{\mbox{\tiny Pl}}^{2}}{2} \int d^{3}x dt N\sqrt{g} \left\{K_{ij}K^{ij}-\lambda K^{2} -g_{0}M_{\mbox{\tiny Pl}}^{2} -g_{1}R-g_{2}M_{\mbox{\tiny Pl}}^{-2}R^{2}-g_{3}M_{\mbox{\tiny Pl}}^{-2}R_{ij}R^{ij}-\right.\\
&\left. \quad -g_{4}M_{\mbox{\tiny Pl}}^{-4}R^{3}-g_{5}M_{\mbox{\tiny Pl}}^{-4}R\left(R^{i}_{\;j}R^{j}_{\;i}\right) -g_{6}M_{\mbox{\tiny Pl}}^{-4}R^{i}_{\;j}R^{j}_{\;k}R^{k}_{\;i} -g_{7}M_{\mbox{\tiny Pl}}^{-4}R\nabla^{2}R -g_{8}M_{\mbox{\tiny Pl}}^{-4}\nabla_{i}R_{jk}\nabla^{i}R^{jk}\right\},
\end{split}\end{equation}

\noindent where $g_{i}$ are coupling constants, $M_{\mbox{\tiny Pl}}$ is the Planck mass and $\nabla_{i}$ denote covariant derivatives. The time coordinate can be rescaled in order to set $g_{1}=-1$, recovering GR value. One also defines the cosmological constant $\Lambda$ as $2\Lambda=g_{0}M_{\mbox{\tiny Pl}}^{2}$. An important feature of the IR limit is the presence of the constant $\lambda$ on the kinetic part of the HL action. GR is recovered provided  $\lambda\rightarrow 1$ (corresponding to the full diffeomorphism invariance), however $\lambda$ must be a running constant, so there is no reason or symmetry that {\it a priori} yields $\lambda=1$ GR value. Phenomenological bounds suggest however that the value of $\lambda$ is quite close to the GR value \cite{Sotiriou:2010wn}.

Performing these redefinitions the HL action reads

\begin{equation}\label{eq:hlaction}\begin{split}
S_{HL}&=\frac{M_{\mbox{\tiny Pl}}^{2}}{2} \int d^{3}x dt N\sqrt{g} \left\{K_{ij}K^{ij}-\lambda K^{2} +R -2\Lambda  -g_{2}M_{\mbox{\tiny Pl}}^{-2}R^{2}-g_{3}M_{\mbox{\tiny Pl}}^{-2}R_{ij}R^{ij}-\right.\\
&\left. \quad g_{4}M_{\mbox{\tiny Pl}}^{-4}R^{3}-g_{5}M_{\mbox{\tiny Pl}}^{-4}R\left(R^{i}_{\;j}R^{j}_{\;i}\right) -g_{6}M_{\mbox{\tiny Pl}}^{-4}R^{i}_{\;j}R^{j}_{\;k}R^{k}_{\;i} -g_{7}M_{\mbox{\tiny Pl}}^{-4}R\nabla^{2}R -g_{8}M_{\mbox{\tiny Pl}}^{-4}\nabla_{i}R_{jk}\nabla^{i}R^{jk}\right\}.
\end{split}\end{equation}

\subsection{Ho\v{r}ava-Lifshitz minisuperspace action}

In order to consistently reduce the number of degrees of freedom when restricting the $3$-metrics of the superspace to be isotropic and homogeneous, one can consider that the restriction is performed directly into the equations of motion, or through the substitution of the RW metric into the Lagrangian density and then obtain the equations of motion for the remaining degrees of freedom. In general, the physical content of these two ways are different, showing that the restriction cannot be done over the Lagrangian unless one properly solves the arising constraints. For the RW metric, without matter fields, these procedures are shown to lead to the same results \cite{Bertolami:1990je}.  Since the HL proposal introduces only an anisotropy between space and time, it does not alter the homogeneity of RW metric and hence the metric (\ref{eq:FLRWmetric}) can be substituted into Eq. (\ref{eq:hlaction}) yield the HL minisuperspace action

\begin{equation}\label{eq:hlminisuperspace2}\begin{split}
S_{HL}&=\frac{M_{\mbox{\tiny Pl}}^{2}\times2\pi^{2}\times 3(3\lambda -1)\sigma^{2}}{2} \int dt N \left\{\frac{-\dot{a}^{2}a}{N^{2}} +\frac{6a}{3(3\lambda-1)} -\frac{2\Lambda\sigma^{2}a^{3}}{3(3\lambda-1)}-\right. \\
&\left.  -M_{\mbox{\tiny Pl}}^{-2}\times\frac{12}{3(3\lambda-1)\sigma^{2}a}\times (3g_{2}+g_{3}) -M_{\mbox{\tiny Pl}}^{-4}\times\frac{24}{3(3\lambda-1)\sigma^{4}a^{3}}\times (9g_{4}+3g_{5}+g_{6})\right\},\end{split}
\end{equation}

\noindent where the spatial integral $\int d^{3}x\sqrt{\gamma}=2\pi^{2}$ has been performed. A further simplification is obtained after choosing units so to satisfy $\sigma^{2}\times6\pi^{2}\times(3\lambda-1)M_{\mbox{\tiny Pl}}^{2}=1$. The minisuperspace action then reads

\begin{equation}\label{eq:hlminisuperspace4}\begin{split}
\!\!\!\!\!\!\!\!\!\!\!\!\!\!\!\!\!S_{HL}&=\frac{1}{2} \int dt N \left\{\frac{-\dot{a}^{2}a}{N^{2}} +\frac{2a}{(3\lambda-1)} -\frac{\Lambda M_{\mbox{\tiny Pl}}^{-2} a^{3}}{18\pi^{2}(3\lambda-1)^{2}} -\frac{24\pi^{2}(3g_{2}+g_{3})}{a} -\right.  \\
&\left.  \frac{288\pi^{4}(3\lambda-1)(9g_{4}+3g_{5}+g_{6})}{a^{3}}\right\}.\end{split}
\end{equation}

Following Ref. \cite{Maeda:2010ke} the dimensionless coupling constants are redefined as

\begin{equation}\label{eq:dimensionlessconstants}
  \begin{split}
    g_{C} =& \frac{2}{3\lambda-1},\\
    g_{\Lambda} =& \frac{\Lambda M_{\mbox{\tiny Pl}}^{-2}}{18\pi^{2}(3\lambda-1)^{2}},\\
    g_{r} =& 24\pi^{2} (3g_{2}+g_{3}), \\  
    g_{s} =& 288\pi^{4}(3\lambda -1) (9g_{4}+3g_{5}+g_{6}).
\end{split}
\end{equation}

\noindent Notice that $g_{C}>0$, which stands for the curvature coupling constant. The sign of $g_{\Lambda}$ follows the sign of the cosmological constant. These two terms are already present in the minisuperspace GR model, but now they depend on $\lambda$. The coupling constants $g_{r}$ and $g_{s}$ can be either positive or negative as their signal does not alter the stability of the HL gravity (cf. Ref. \cite{Maeda:2010ke}). As discussed in Ref. \cite{Sotiriou:2009bx}, physically, $g_{r}$ corresponds to the coupling constant for the term behaving as a radiaton and $g_{s}$ stands for the term behaving as  ``stiff'' matter ($p=\rho$ equation of state). The minisuperspace action is finally written as \cite{Sotiriou:2009bx}

\begin{equation}\label{eq:hlminisuperspacefinal}
S_{HL}=\frac{1}{2}\int dt \left(\frac{N}{a}\right)\left[-\left(\frac{a}{N}\dot{a}\right)^{2} +g_{C}a^{2} -g_{\Lambda}a^{4}-g_{r}-\frac{g_{s}}{a^{2}} \right].
\end{equation}

\subsection{Ho\v{r}ava-Lifshitz minisuperspace Hamiltonian and Wheeler-DeWitt equation}

The canonical conjugate momentum associated to $a$ is obtained using Eq. (\ref{eq:hlminisuperspacefinal})

\begin{equation}\label{eq:canonicalmomentum}
 \Pi_{a}=\frac{\partial \mathcal{L}}{\partial \dot{a}}=-\frac{a}{N}\dot{a}.
\end{equation}

The Ho\v{r}ava-Lifshitz minisuperspace Hamiltonian density is performed using Eqs. (\ref{eq:hlminisuperspacefinal}) and (\ref{eq:canonicalmomentum}) 

\begin{equation}\label{eq:hamiltonian}
 H=\Pi_{a}\dot{a}-\mathcal{L}=\frac{1}{2}\frac{N}{a}\left(-\Pi_{a}^{2}-g_{C}a^{2} +g_{\Lambda}a^{4}+g_{r}+\frac{g_{s}}{a^{2}}\right).
\end{equation}

The next step to implement the quantum cosmology programme involves promoting the classical minisuperspace Hamiltonian into an operator on which the so-called wave function of the universe is applied to \cite{DeWitt:1967yk,Hartle:1983ai}. 

This is subtle point in HL gravity since there is no global diffeomorphism, just a foliation-preserving diffeomorphism \cite{Horava:2009uw}. This can be also seen as the lapse function no longer depends on the space-time variables, as in GR, but now it depends only on the global time $N=N(t)$, as discussed in Sec. \ref{sec:Introduction}. This implies that the Hamiltonian constraint is not local, however this problem can be circumvented for an homogeneous metric, like Eq. (\ref{eq:FLRWmetric}), as the integration over space can be performed as seen above. The canonical quantization is obtained by promoting the canonical conjugate momentum into an operator, $i.e.$ $\Pi_{a}\mapsto -i\frac{d}{d a}$. Due to ambiguities in the operator ordering, one chooses $\Pi_{a}^{2}=-\frac{1}{a^{p}}\frac{d}{d a}\left(a^{p}\frac{d}{d a}\right)$ \cite{Hartle:1983ai}. The resulting WdW equation is then obtained

 \begin{equation}\label{eq:wdwequationp}
  \left\{\frac{1}{a^{p}}\frac{d}{d a}\left(a^{p}\frac{d}{d a}\right)-g_{C}a^{2} +g_{\Lambda}a^{4}+g_{r}+\frac{g_{s}}{a^{2}}\right\}\Psi(a)=0.
 \end{equation}

The choice of $p$ does not modify the semiclassical analysis \cite{KolbTurner:1989}, hence one chooses $p=0$, and the WdW equation is written as

 \begin{equation}\label{eq:wdwequation}
  \left\{\frac{d^{2}}{d a^{2}}-g_{C}a^{2} +g_{\Lambda}a^{4}+g_{r}+\frac{g_{s}}{a^{2}}\right\}\Psi(a)=0.
 \end{equation}

This equation is similar to the one-dimensional Schr\"{o}dinger equation for $\hbar=1$ and a particle with $m=1/2$  with $E=0$ and potential 

\begin{equation}\label{eq:HLpotential}
V(a)=g_{C}a^{2}-g_{\Lambda}a^{4}-g_{r}-\frac{g_{s}}{a^{2}}.
\end{equation}

\subsection{Ho\v{r}ava-Lifshitz minisuperspace potentials}

The WdW equation derived in the last section resembles an unidimensional Schr\"{o}dinger equation with potential given by Eq. (\ref{eq:HLpotential}). Classically the allowed regions are such that $V(a)\leq 0$ since $E=0$. A complete analysis of the phase structure of HL FLRW cosmologies was performed in Ref. \cite{Maeda:2010ke}. 

Notice that the first two terms of Eq. (\ref{eq:HLpotential})  are the usual GR terms of the quantum cosmology analysis \cite{Hartle:1983ai}. HL gravity introduces the last two terms which dominate for $a\ll 1$, $i.e.$ are relevant at short distances, presumably at the very early universe, where the GR description must be replaced by the quantum gravity one. At the very early universe, this potential is dominated by the term $-g_{s}/a^{2}$, implying that for $g_{s}<0$, this potential exhibits a ``barrier'' that might prevent space-time to get singular. The  case $g_ {s}>0$ is not examined as it leads to a  cosmology that cannot be suitably  investigated using QC techniques. Notice that the detailed balance condition yields $g_{s}=0$. 

The choice $g_{s}<0$ splits the discussion into three distinct scenarios, for positive, negative and vanishing cosmological constant. In what follows one studies the cases ensued by these choices for the coefficients Eqs. (\ref{eq:dimensionlessconstants}).

\subsubsection{$\Lambda=0$ case}
\begin{figure}[htb]
\centering 
\includegraphics[width=0.4\textwidth]{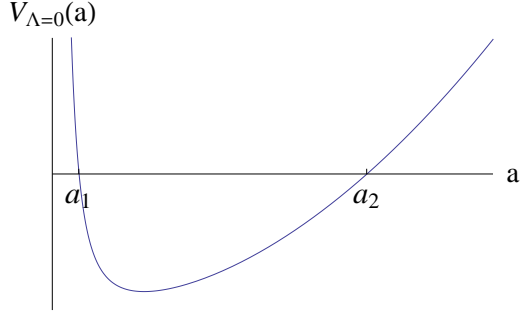} 
\caption{Potential for $\Lambda=0$.} \label{fig:NullLambdapotential}
\end{figure}

In this case the curvature term dominates at large distances. The universe oscillates between $a_{1}$ and $a_{2}$. The potential, depicted in Figure \ref{fig:NullLambdapotential}, is written as

\begin{equation}\label{eq:NullLambdaPotential}
V_{\Lambda=0}(a)=g_{C}a^{2}-g_{r}-\frac{g_{s}}{a^{2}}. 
\end{equation}

\subsubsection{$\Lambda \neq0$ case}
For large $a$, the potential is dominated by the cosmological term, and given by

\begin{equation}\label{eq:NonullLambdaPotential}
V_{\Lambda\neq0}(a)=g_{C}a^{2}-g_{\Lambda}a^{4}-g_{r}-\frac{g_{s}}{a^{2}}. 
\end{equation}

\begin{figure}
  \centering
\subfloat[$\Lambda > 0$ Potential]{\includegraphics[width=0.4\textwidth]{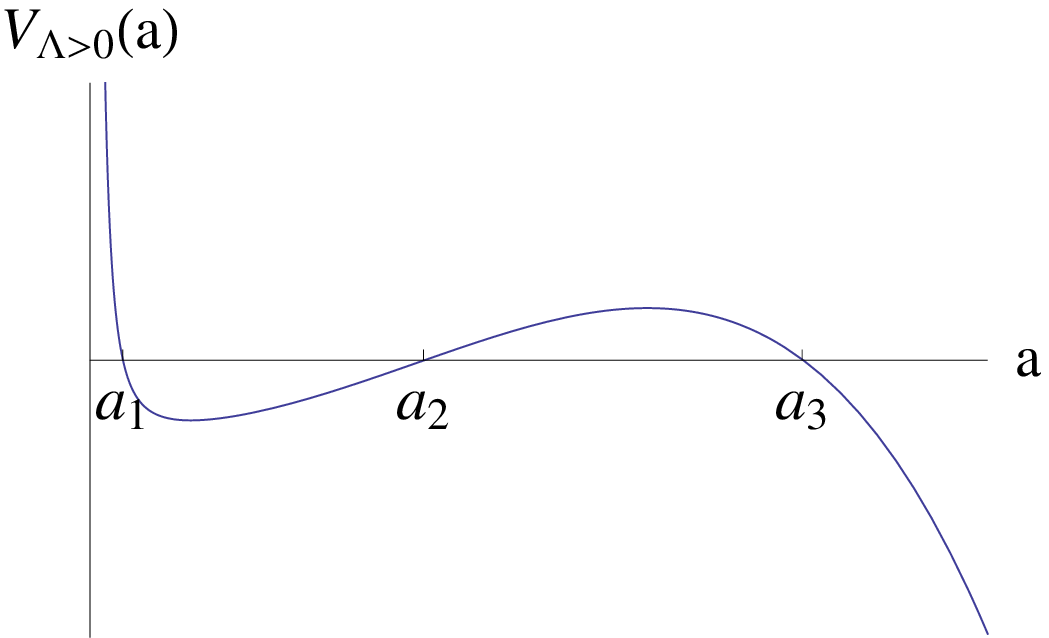}\label{fig:PositiveLambda}} 
\hspace{1cm} 
\subfloat[$\Lambda < 0$ Potential]{\includegraphics[width=0.4\textwidth]{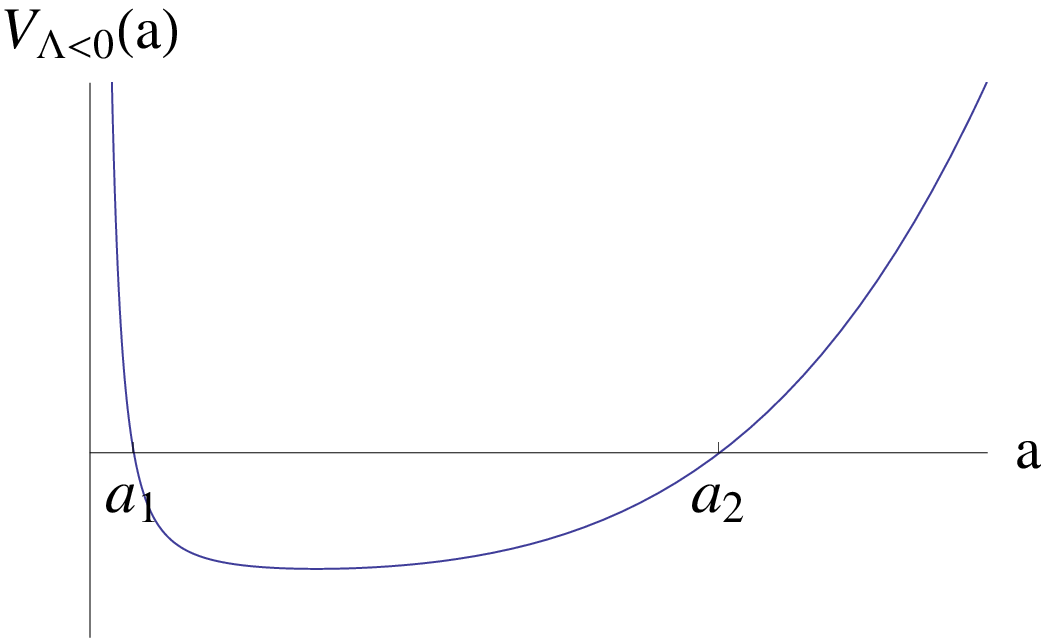}\label{fig:NegativeLambda}}
\caption{Potentials for non-vanishing cosmological constant.} \label{fig:nonvanishingcosmologicalconstantpotential}
\end{figure}

The sign of $g_{\Lambda}$ follows the sign of the cosmological constant $\Lambda$. For positive $\Lambda$ the potential is depicted in Figure \ref{fig:PositiveLambda} and the negative $\Lambda$ case in Figure \ref{fig:NegativeLambda}. For a positive cosmological constant, one considers a potential that has three positive roots  ($a_{1}$, $a_{2}$ and $a_{3}$) hence  there are two classically allowed regions for $a_{1}<a<a_{2}$ and $a_{3}<a$ and a forbidden region where $a_{2}<a<a_{3}$. There is another possibility, discussed in Refs. \cite{Mukohyama:2010xz,Maeda:2010ke}, in which the potential has only one real positive root, namely $a_ {1}$. The expression for the roots of Eq. (\ref{eq:NonullLambdaPotential}), can be found in Ref. \cite{Abramowitz:1980}, and will not be presented here as their expresions will not play any role in what follows. The potential can be factorized as

\begin{equation}\label{eq:PositiveLambdaPotentialFactorized}
V_{\Lambda > 0}(a)=-\frac{g_{\Lambda}}{a^2}\left(a^{2}-a^{2}_{1}\right)\left(a^{2}-a^{2}_{2}\right)\left(a^{2}-a^{2}_{3}\right). 
\end{equation}

By the same token, for a negative cosmological constant, one finds  a similar behaviour already present in the $\Lambda=0$ case: classically, the universe oscillates between $a_{1}$ and $a_{2}$. This implies that the potential Eq. (\ref{eq:NonullLambdaPotential}) reads

\begin{equation}\label{eq:NegativeLambdaPotentialFactorized}
V_{\Lambda < 0}(a)=-\frac{g_{\Lambda}}{a^2}\left(a^{2}-a^{2}_{1}\right)\left(a^{2}-a^{2}_{2}\right)\left(a^{2}+a^{2}_{i}\right), 
\end{equation}

\noindent where $a=\pm i a_{i}$ are  the imaginary roots of this potential and $a_{i}$ is real.

\section{Solutions of the Wheeler-DeWitt equation}\label{sec:WdWEquationSolutions}

Having described the three types of potentials one encounters, the task is to solve the WdW equation (\ref{eq:wdwequation}). If $\Lambda\neq0$, the cosmological constant term is quartic so the Eq. (\ref{eq:wdwequation}) cannot be solved in a closed form, and  the WKB approximation will be employed.

\subsection{Boundary Conditions}

To find suitable boundary conditions for the WdW equation one has to rely on  physical assumptions. The most discussed choices are the de Witt boundary condition \cite{DeWitt:1967yk}, the ``no-boundary" proposal \cite{Hartle:1983ai,Halliwell:2009rb} and the tunnelling boundary condition \cite{Vilenkin:1983xq}.

The deWitt boundary condition \cite{DeWitt:1967yk} is the one in which the wave function of the universe is required to vanish wherever there is a classical singularity. It is inspired on quantum mechanics, and it is suitable for the cases under study here, as there is a potential barrier (bounce) for $a\ll1$ yielding that the singularity is inside a classically forbidden region (cf. Ref. \cite{Kiefer:2010zz}). The de Witt boundary condition is expressed, for FLRW models as: 

\begin{equation}\label{eq:dWBC}
\psi_{dW}(a=0)=0.
\end{equation}

The ``no-boundary" condition, of Hartle and Hawking \cite{Hartle:1983ai}, arises from obtained using the Euclidean path integral formalism. In that formalism, the ground state for the wave function of the universe is written as (cf. \cite{Bertolami:1998hm})

\begin{equation} \label{eq:HartleHawingWF}
\psi(a)=\int_{C} \left[da\right] \exp{\left(-I\right)},
\end{equation}

\noindent where $C$ denotes that the integral is taken over compact manifolds, and $I$ is the Euclidean version of the action defined in Sec. \ref{sec:WdWEquation} such that the corresponding Euclidean action $I=-iS_{HL}$ can be obtained from Eq. (\ref{eq:hlminisuperspacefinal}) using $d\tau=iN dt$ and the $N=1$ gauge:

\begin{equation}\label{eq:hlminisuperspaceeuclidean}\begin{split}
I&=-iS_{HL}=\frac{1}{2}\int d\tau \left[-a\left(\frac{da}{d\tau}\right)^{2} -g_{C}a +g_{\Lambda}a^{3}+\frac{g_{r}}{a}+\frac{g_{s}}{a^{3}} \right],
\end{split}
\end{equation}

\noindent where $\tau$ is the Euclidean time. It is possible to evaluate $\psi(a)$ nearby $a=0$ \cite{Bertolami:1998hm,Bertolami:1996pc}. In this case, it can be proved that for $\tau\ll1$ (close to $a=0$), one has $\frac{da}{d\tau}=1$ \cite{Halliwell:2009rb}. Substituting these conditions into the Euclidean version of action Eq. (\ref{eq:hlminisuperspaceeuclidean}), and integrating from $0$ to $\Delta \tau$, one finds 

\begin{equation}
I=\frac{1}{2}\int d\tau \left[-(1+g_{C})\tau +g_{\Lambda}\tau^{3}+\frac{g_{r}}{\tau}+\frac{g_{s}}{\tau^{3}} \right],
\end{equation}

\noindent which is $I\rightarrow -\infty$, for $g_{r}\neq0$ and $g_{s}\neq0$ yielding a divergent wave function. This shows that the ``no-boundary'' condition is not suitable to the problem under study.

It is important to notice that the boundary condition $\psi(0)=0$ does not mean that there is a quantum avoidance of the classical singularity given that it is a sufficient but not a necessary condition \cite{Kiefer:2010zz,Kamenshchik:2007zj,Bastos:2009kp}. In the above references, some examples are given where $\psi\rightarrow0$, but $\int da |\psi(a)|^{2}$ diverges, and conversely cases where $\int da |\psi(a)|^{2}\rightarrow0$, but $\psi$ diverges. The conditions under which the classical singularity is removed or avoided by quantum mechanics are understood only in specific cases (cf. \cite{Kiefer:2010zz,Bastos:2009kp} and references therein).

\subsection{WdW equation solution for $a\ll 1$}

This region corresponds to the very early universe, where the HL terms dominate. This HL epoch is expected since any theory of quantum gravity is supposed to alter the GR description of the  structure of the space-time at small distances. For this case, the WdW Eq.  (\ref{eq:wdwequation}) reads

\begin{equation}
\left\{\frac{d^{2}}{da^{2}}+\frac{g_{s}}{a^{2}}\right\}\psi(a)=0,
\end{equation}

\noindent which is an Euler equation whose solution is $\psi(a)=a^{\delta}$ and $\delta=\frac{1}{2}\pm\frac{1}{2}\sqrt{1-4g_s}$. As the solution must be real and $\psi(0)=0$, one finds that the wave function for $a\ll1$ goes as

\begin{equation}\label{eq:HLdominantsolution}
\psi(a)\sim a^{\frac{1}{2}+\frac{1}{2}\sqrt{1-4g_s}},
\end{equation}

\noindent which yields that $g_s\leq1/4$, a ``quantum'' bound for $g_ s$. Notice that this coefficient is unconstrained by classical consideration \cite{Maeda:2010ke}. Although the potential for $g_{s}>0$ corresponds to an infinite well, this quantum bound gives rise to a mild singularity, which admits a well-defined mathematical treatment (cf. Ref. \cite{Voronov:2006}).

\subsection{WdW equation solutions for $a\gg1$}

This limit corresponds to the very late universe, which is dominated by the curvature and cosmological constant terms. These terms are already present in the usual GR quantum cosmology setup \cite{Hartle:1983ai,Halliwell:2009rb}, reflecting the fact that GR behaviour is recovered at large distances. For $\Lambda=0$, Eq. (\ref{eq:wdwequation}) for $a\gg1$ is given by

\begin{equation}
\left\{\frac{d^{2}}{da^{2}}-g_{C}a^{2}\right\}\psi(a)=0,
\end{equation}

\noindent which has the following asymptotic solution

\begin{equation}\label{eq:curvaturedominantsolution}
\psi(a)\sim e^{-\frac{\sqrt{g_{C}}}{2} a^{2}}.
\end{equation}

\noindent Thus, as expected, the wave function has an exponential behaviour, since $a\gg1$ corresponds to a classically forbidden region for the potential Eq. (\ref{eq:NullLambdaPotential}). 

For the positive cosmological constant case ($g_{\Lambda}>0$), the WdW Eq. (\ref{eq:wdwequation})  reads

\begin{equation}
\left\{\frac{d^{2}}{da^{2}}+g_{\Lambda}a^{4}\right\}\psi(a)=0,
\end{equation}

\noindent   whose  asymptotic solution is given by a combination of Bessel and Neumann functions with $\nu=1/6$ (cf. Eq. $8.491.7$ of Ref. \cite{Gradshteyn:1994}). Since the limit $a\gg1$  is being considered and the Neumann functions only diverge at $a=0$, these two functions are admissible:

\begin{equation}\label{eq:psipositivelambda}
\psi(a)=\bar{C_{1}}\sqrt{a}J_{\frac{1}{6}}\left(\frac{\sqrt{g_{\Lambda}}}{3}a^{3}\right) + \bar{C_{2}}\sqrt{a}N_{\frac{1}{6}}\left(\frac{\sqrt{g_{\Lambda}}}{3}a^{3}\right),
\end{equation}

\noindent where $\bar{C_{1}}$ and $\bar{C_{2}}$ are constants. A further analysis show that the asymptotic expansions for  Bessel and Neumann functions of any  order ($\nu$) and large arguments ($|z|\rightarrow\infty$) is given by (cf.  Eqs. $9.2.1$ and $9.2.2$ of Ref. \cite{Abramowitz:1980}):

\begin{equation}
  \begin{split}
    J_{\nu} (z)\approx& \sqrt{\frac{2}{\pi z}}\cos\left(z-\frac{\nu\pi}{2}-\frac{\pi}{4}\right),\\
    N_{\nu} (z)\approx& \sqrt{\frac{2}{\pi z}}\sin\left(z-\frac{\nu\pi}{2}-\frac{\pi}{4}\right).
\end{split}
\end{equation}

Substituting these asymptotic expressions into Eq. (\ref{eq:psipositivelambda}) one finds the asymptotic behaviour of the wave function for large $a$:

\begin{equation}\label{eq:psipositivelambdafinal}
\psi(a)=\frac{C_{1}}{a}\cos\left(\frac{\sqrt{g_{\Lambda}}}{3}a^{3} -\frac{\pi}{12}-\frac{\pi}{4}\right) + \frac{C_{2}}{a}\sin\left(\frac{\sqrt{g_{\Lambda}}}{3}a^{3} -\frac{\pi}{12}-\frac{\pi}{4}\right),
\end{equation}

\noindent where $C_{i}=\bar{C_{i}} \sqrt{\frac{6}{\pi \sqrt{g_{\Lambda}} }}$, for $i=1,2$. Notice that this wave function is oscillatory, denoting that this is a classically allowed region and damped as $|\psi(a)|^{2}\sim a^{-2}$. Not surprisingly, the same behaviour is found in Ref. \cite{Hartle:1983ai} when the cosmological constant dominates the evolution of the universe, and the GR regime is recovered. This issue will be discussed in Sec. \ref{sec:interpretation}. Interestingly, the WKB method yields the same asymptotic expression given by Eq. (\ref{eq:psipositivelambda}) \cite{Bastos:2009kp,Voronov:2006}.

For $\Lambda<0\Rightarrow g_{\Lambda}<0$, Eq. (\ref{eq:wdwequation})  is written as

\begin{equation}
\left\{\frac{d^{2}}{da^{2}}-\left(-g_{\Lambda}\right)a^{4}\right\}\psi(a)=0,
\end{equation}

\noindent whose asymptotic solution is  a combination of the modified Bessel functions, $I_{\nu}(z)$ and $K_{\nu}(z)$ (cf. $8.406$ of Ref.  \cite{Gradshteyn:1994}) of order $\nu=1/6$. However, $I_{\nu}(z)$ grows exponentially as $z\rightarrow\infty$ (cf. Eq. (\ref{eq:asymptoticmodifiedbessel})), hence only $K_{\nu}(z)$ represents an acceptable solution for large $a$. The wave function in that limit is given by

\begin{equation}\label{eq:psinegativelambda}
\psi(a) \sim \sqrt{a}K_{\frac{1}{6}}\left(\frac{\sqrt{-g_{\Lambda}}}{3}a^{3}\right).
\end{equation}

\noindent As above, asymptotic expansions for  modified Bessel functions of any order ($\nu$) and large arguments ($|z|\rightarrow\infty$) are obtained using Eqs. $9.7.1$ and $9.7.2$ of Ref. \cite{Abramowitz:1980}: 

\begin{equation}\label{eq:asymptoticmodifiedbessel}
\begin{split}
I_{\nu} (z)\approx& \frac{e^{z}}{\sqrt{2\pi z}},\\
K_{\nu} (z)\approx& \sqrt{\frac{\pi}{2 z}}e^{-z}.
\end{split}
\end{equation}

One then gets that the wave function for the very late universe

\begin{equation}\label{eq:psinegativelambdafinal}
\psi(a)\sim\frac{1}{a}e^{\frac{-\sqrt{-g_{\Lambda}}}{3}a^{3}}.
\end{equation}

\noindent Thus, one concludes that for $a\gg1$ the wave function is, as expected, strongly suppressed in this limit, given that this region is classically forbidden.

\subsection{WdW equation solution for  $\Lambda=0$}

If the cosmological constant vanishes the WdW Eq. (\ref{eq:wdwequation}) reads\footnote{In a quantum mechanical context, this equation was solved in the Problem 4 of \textsection36 of Ref. \cite{Landau:1980}.}

 \begin{equation}\label{eq:wdwequationnullcosmologicalconstant}
  \left\{\frac{d^{2}}{d a^{2}}-g_{C}a^{2} +g_{r}+\frac{g_{s}}{a^{2}}\right\}\Psi(a)=0.
 \end{equation}

After a change of variables, $x=g_{C}^{1/4}a$, Eq. (\ref{eq:wdwequationnullcosmologicalconstant}) reads

\begin{equation}\label{eq:wdwequationnullcosmologicalconstantmodified}
  \left\{\frac{d^{2}}{d x^{2}}-x^{2} +\frac{g_{r}}{g_{C}^{1/2}}+\frac{g_{s}}{x^{2}}\right\}\Psi(x)=0.
 \end{equation}

This equation can be exactly solved in terms of the associate Laguerre functions. Indeed,  the following differential equation (cf. Eq. $22.6.18$ of Ref. \cite{Abramowitz:1980})

\begin{equation}\label{eq:abramowitzlaguerre}
  \left\{\frac{d^{2}}{d x^{2}} + 4n +2\alpha + 2 -x^{2} +\frac{1-4\alpha^{2}}{4x^{2}}\right\}y(x)=0,
 \end{equation}
 
 \noindent has as solution $y(x)=e^{-\frac{x^{2}}{2}}x^{\alpha)+\frac{1}{2}}L_{n}^{(\alpha)}(x^{2})$. Here $n$ is a positive integer, and $L_{n}^{(\alpha)}$ are  associate Laguerre functions.  Comparing Eqs. (\ref{eq:wdwequationnullcosmologicalconstantmodified}) and (\ref{eq:abramowitzlaguerre}), one finds that
 
 \begin{equation}\label{eq:quantizationcondition}
  \begin{split}
    \alpha = & \frac{1}{2}\sqrt{1-4g_{s}},\\
    \frac{g_{r}}{\sqrt{g_{C}}} =& 4n + 2 + 2\alpha, \\
    \psi(a)=&Ne^{-\frac{\sqrt{g_{C}}a^{2}}{2}}\left(g_{C}^{1/4}a\right)^{\alpha+\frac{1}{2}}L_{n}^{(\alpha)}(\sqrt{g_{C}}a^{2})
\end{split}
\end{equation}

\noindent where $N$ is a normalization constant to be obtained below. As $g_{s}<0$, $\alpha>0$, the wave function $\psi(0)$ is regular. Comparing with Eq. (\ref{eq:HLdominantsolution}), one also finds that the ratio $g_{r}/\sqrt{g_{C}}$ must be quantized. This is not surprising given that one is solving the Schr\"{o}dinger equation with $E=0$, which for a bounded potential has a discrete spectrum. Thus, $E=0$ is an eigenvalue only for specific values of the coefficients and these values must be quantized. Another interesting feature of this solution is that it is normalizable. Indeed, using Eq. (\ref{eq:quantizationcondition}) and Eq. $8.980$ of Ref. \cite{Gradshteyn:1994}, one obtains for the normalization condition that $N=\sqrt{\frac{2n! g_{C}^{1/4}}{\Gamma(n+\alpha+1)}}$. The complete solution,  for the $\Lambda=0$ WdW equation satisfying the $\psi(0)=0$ boundary condition is given by

\begin{equation}\label{eq:WdWWFSolutionNullLambda}
\psi(a)=\sqrt{\frac{2n! g_{C}^{1/4}}{\Gamma(n+\alpha+1)}}e^{-\frac{\sqrt{g_{C}}a^{2}}{2}}\left(g_{C}^{1/4}a\right)^{\alpha+\frac{1}{2}}L_{n}^{(\alpha)}(\sqrt{g_{C}}a^{2}).
\end{equation}	

\begin{figure}
  \centering
\subfloat[$n = 0$]{\includegraphics[width=0.4\textwidth]{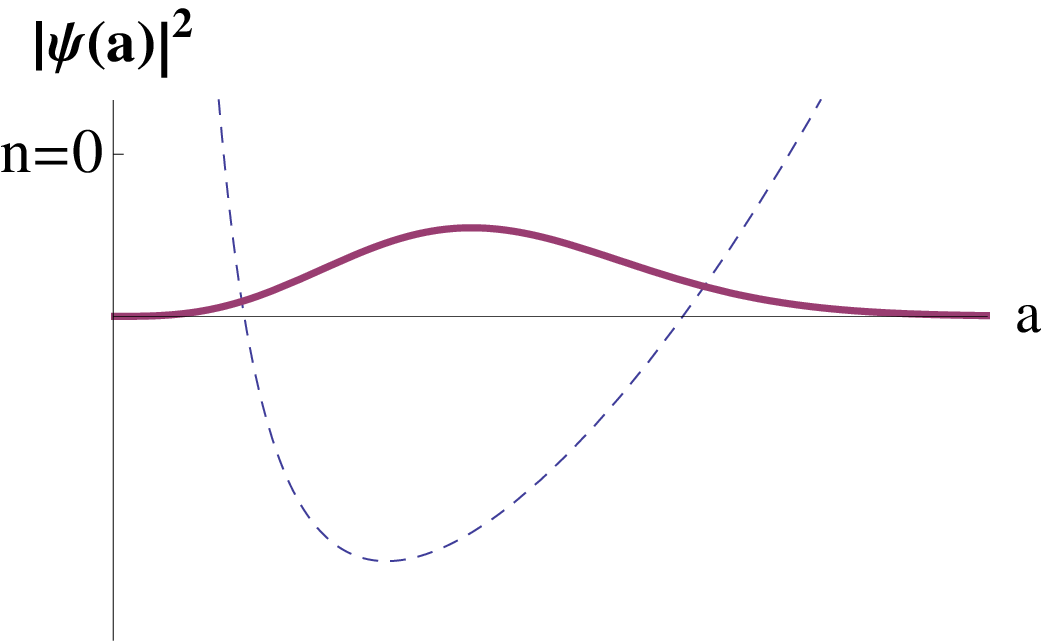}\label{fig:N0psisquare}} 
\hspace{0.5cm} 
\subfloat[$n = 1$]{\includegraphics[width=0.4\textwidth]{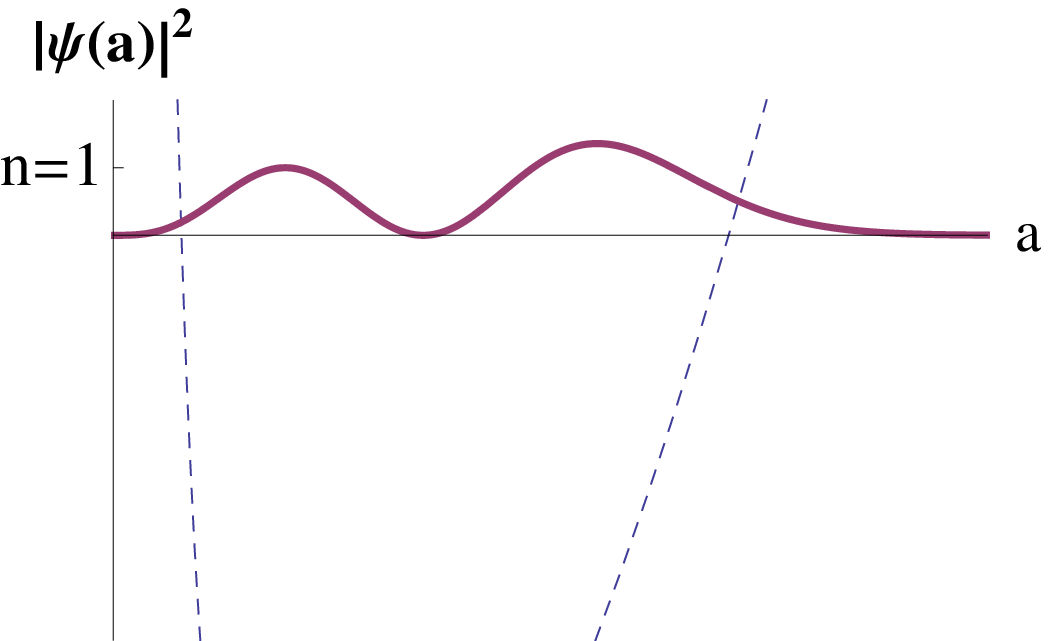}\label{fig:N1psisquare}}
\vspace{0.5cm} 
\subfloat[$n = 2$]{\includegraphics[width=0.4\textwidth]{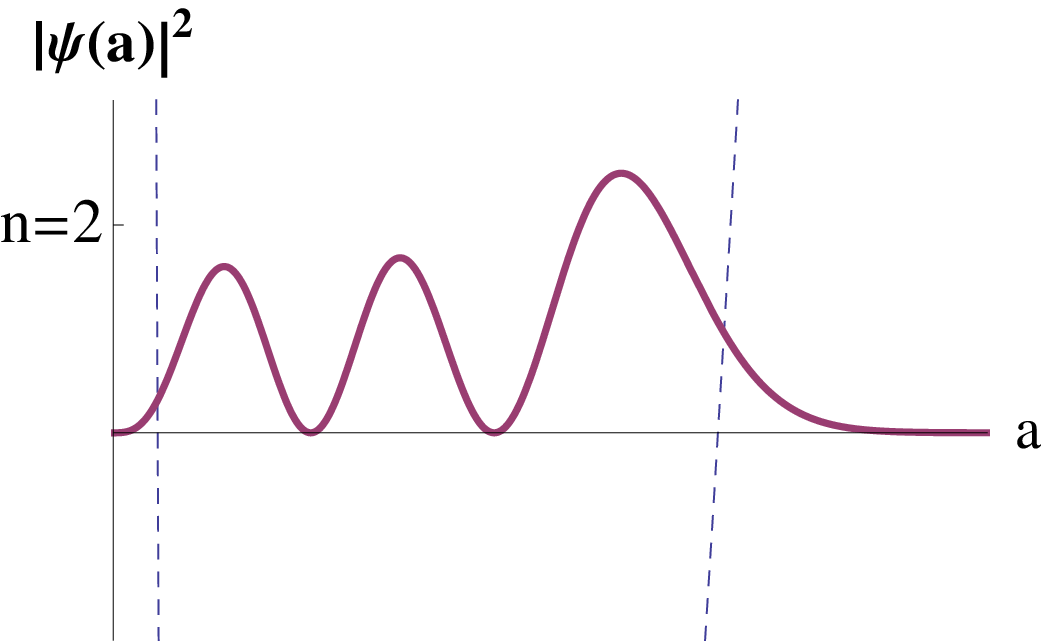}\label{fig:N2psisquare}} 
\hspace{0.5cm} 
\subfloat[$n = 10$]{\includegraphics[width=0.4\textwidth]{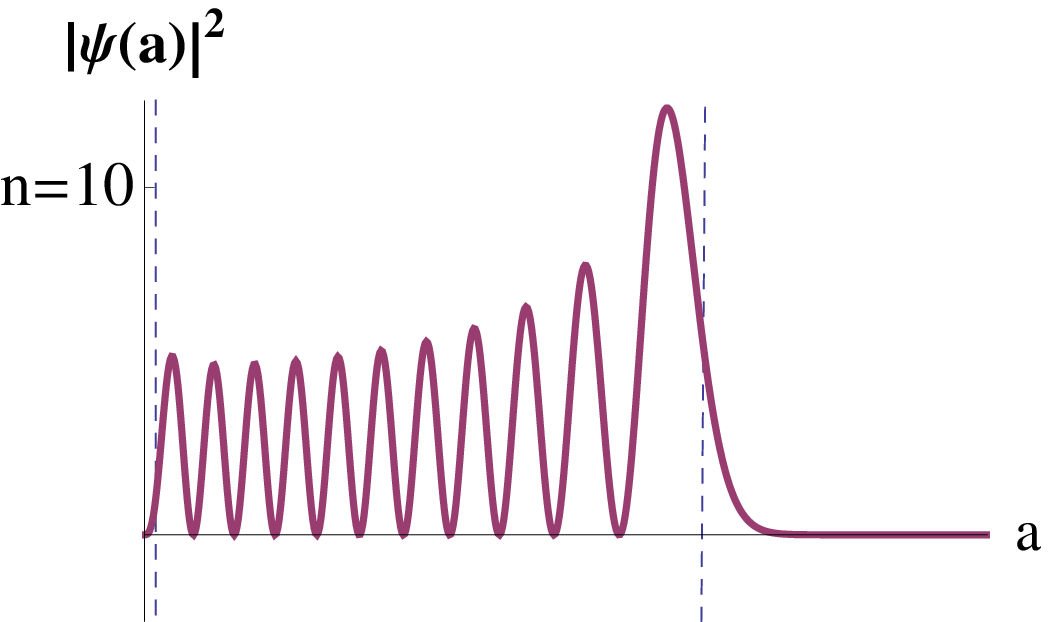}\label{fig:N10psisquare}}
\caption{Probability density for the wave function of the universe for diferent values of $n$. The dashed line plot represents the potential Eq. (\ref{eq:NullLambdaPotential}). } \label{fig:probabilitydensity}
\end{figure}

This wave function behaves as Eq. (\ref{eq:HLdominantsolution}) for $a\ll1$ and as Eq. (\ref{eq:curvaturedominantsolution}) for $a\gg1$. The Laguerre associated function $L_{n}^{\alpha}$ is an $n$-th order polynomial which yields that the wave function has $n$ nodes. It is not difficult to verify that for a fixed $g_{s}$, any $g_r$ obtained through the quantization condition Eq. (\ref{eq:quantizationcondition}) implies that the potential Eq. (\ref{eq:NullLambdaPotential}) has two positive real roots and so the universe is oscillating whatever value of $n\geq0$ is chosen.  For large $g_{r}$ values,  which implies that $n$ is large, the very structure of the wave function shows that the universe is almost classical (this can be also seen by inspection of the probability density plots for large $n$ as shown in Figure \ref{fig:N10psisquare}). This result can be understood in terms of Bohr's correspondence principle, according to which the classical behaviour is obtained from the quantum one in the limit of large quantum numbers. The probability density distribution for the wave function of the universe, $|\psi(a)|^{2}$, for some $n$ values are plotted in Figure \ref{fig:probabilitydensity}. One  clearly sees that the solution is highly suppressed in the classically forbidden region, and it is oscillating with $n$ nodes in the classically allowed region. Finally, it is straightforward to show that the singularity is avoided in this model given that the probability to find the universe at $a=0$ vanishes due to the HL gravity terms. 

\subsection{WdW solution for $\Lambda>0$}

If $g_{\Lambda}\neq0$, Eq. (\ref{eq:wdwequation}) cannot be analitically solved. The behaviour of the wave function for large $a$ and nearby the singularity $a=0$ were already discussed. For the intermediate regions where the curvature term starts to become relevant, after the HL epoch (very early universe), one has to rely on the WKB approximation, which for the classically allowed region is given by \cite{Landau:1980}

\begin{equation}\label{eq:allowedWKBwavefunction}
\psi(a)\approx\frac{1}{|V(a)|^{1/4}}\exp\left[\pm i \int_{a_{1}}^{a} \sqrt{|V(a)|} da \right],
\end{equation}

\noindent where $\pm$ symbol denotes that one must consider a combination of these two exponentials, $V(a)$ is the potential Eq. (\ref{eq:PositiveLambdaPotentialFactorized}) and $a_{1}$ is the classical turning point, $i.e.$ $V(a_{1})=0$. The following integral must be solved

\begin{equation}
\int_{a_{1}}^{a} \sqrt{|V(a)|} da = \sqrt{g_{\Lambda}}\int_{a_{1}}^{a} \frac{\sqrt{(a^{2}-a_{1}^{2})(a_{2}^{2}-a^{2})(a_{3}^{2}-a^{2})}}{a} da.
\end{equation}

\noindent One uses that $V(a)\leq0$ for the classically allowed region and, hence $|V(a)|=-V(a)$. This integral is valid for $a_{1}<a<a_{2}<a_{3}$. Changing the variables to $t=a^{2}$ and rationalizing the square root, one finds

\begin{equation}
\int_{a_{1}}^{a} \sqrt{|V(a)|} da = \frac{\sqrt{g_{\Lambda}}}{2}\int_{a_{1}^{2}}^{a^{2}} \frac{(t-a_{1}^{2})(a_{2}^{2}-t)(a_{3}^{2}-t)}{t\sqrt{(t-a_{1}^{2})(a_{2}^{2}-t)(a_{3}^{2}-t)}} dt.
\end{equation}

This integral can be written as a sum of elliptic integrals \cite{Gradshteyn:1994,Byrd:1954}. Using the formulas $3.131.3$, $3.132.2$, $3.137.3$ of Ref. \cite{Gradshteyn:1994} and the reduction formula $230.01$ of Ref. \cite{Byrd:1954} one finds after a rather long although straightforward computation

\begin{equation}\label{eq:WKBallowedargument}\begin{split}
\int_{a_{1}}^{a} \sqrt{|V(a)|} da =& \frac{\sqrt{g_{\Lambda}}}{3}\left\{\sqrt{(a^{2}-a_{1}^{2})(a_{2}^{2}-a^{2})(a_{3}^{2}-a^{2})} -\left(a_{1}^{2}+a_{2}^{2}+a_{3}^{2}\right)E(\gamma,q) +\right. \\
 & \left. \left(\frac{2a_{1}^{2}a_{2}^{2}+a_{1}^{2}a_{3}^{2}+a_{2}^{2}a_{3}^{2}-a_{3}^{4}}{\sqrt{a_{3}^{2}-a_{1}^{2}}}\right)F(\gamma,q)-\frac{3a_{2}^{2}a_{3}^{2}}{\sqrt{a_{3}^{2}-a_{1}^{2}}} \;\Pi\left(\gamma,\frac{a_{1}^{2}-a_{2}^{2}}{a_{1}^{2}},q\right)\right\},
\end{split}
\end{equation}

\noindent where $\gamma=\arcsin \sqrt{\frac{a^{2}-a_{1}^{2}}{a_{2}^{2}-a_{1}^{2}}}$, $q=\sqrt{\frac{a_{2}^{2}-a_{1}^{2}}{a_{3}^{2}-a_{1}^{2}}}$. $F(\varphi,k)$ is an elliptic integral of first kind, $E(\varphi,k)$ is an elliptic integral of second kind and  $\Pi\left(\varphi,n,k\right)$ is the elliptic integral of the third kind (cf. \cite{Gradshteyn:1994}). Substituting Eqs. (\ref{eq:PositiveLambdaPotentialFactorized}) and (\ref{eq:WKBallowedargument}) into Eq. (\ref{eq:allowedWKBwavefunction}) one finds the WKB wave function of the universe. This approximation is valid only if $\frac{dV(a)}{da}\ll |V(a)|^{3/2}$ \cite{Landau:1980}.

For a positive cosmological constant, the classically forbidden region, $a_{1}<a_{2}<a<a_{3}$, has the following WKB wave function 

\begin{equation}\label{eq:forbiddenWKBwavefunction}
\psi(a)=\frac{C_{1}}{|V(a)|^{1/4}}\exp\left[ \int_{a_{2}}^{a} \sqrt{|V(a)|} da \right]+\frac{C_{2}}{|V(a)|^{1/4}}\exp\left[ -\int_{a_{2}}^{a} \sqrt{|V(a)|} da \right],
\end{equation}

\noindent and one has to solve 

\begin{equation}
\int_{a_{2}}^{a} \sqrt{V(a)} da = \sqrt{g_{\Lambda}}\int_{a_{2}}^{a} \frac{\sqrt{(a^{2}-a_{1}^{2})(a^{2}-a_{2}^{2})(a_{3}^{2}-a^{2})}}{a} da.
\end{equation}

Following the same steps as before, this integral can be written as

\begin{equation}
\int_{a_{2}}^{a} \sqrt{V(a)} da = \frac{\sqrt{g_{\Lambda}}}{2}\int_{a_{2}^{2}}^{a^{2}} \frac{(t-a_{1}^{2})(t-a_{2}^{2})(a_{3}^{2}-t)}{t\sqrt{(t-a_{1}^{2})(t-a_{2}^{2})(a_{3}^{2}-t)}} dt,
\end{equation}

\noindent which can be solved using the formulas $3.131.5$, $3.132.4$, $3.137.5$ of \cite{Gradshteyn:1994} and the reduction formula $230.01$ of \cite{Byrd:1954}. One gets

\begin{equation}\label{eq:WKBforbiddenargument}\begin{split}
\int_{a_{2}}^{a} \sqrt{V(a)} da =& \frac{\sqrt{g_{\Lambda}}}{3}\left\{\sqrt{(a^{2}-a_{1}^{2})(a^{2}-a_{2}^{2})(a_{3}^{2}-a^{2})} -\frac{(a_{1}^{2}+a_{2}^{2}+a_{3}^{2})(a_{2}^{2}-a_{1}^{2})}{\sqrt{a_{3}^{2}-a_{1}^{2}}}\;\Pi\left(\kappa,p^{2},p\right)\right. \\
 +& \left. \left(\frac{a_{1}^{4}-a_{1}^{2}a_{2}^{2}-a_{1}^{2}a_{3}^{2}+a_{2}^{2}a_{3}^{2}}{\sqrt{a_{3}^{2}-a_{1}^{2}}}\right)F(\kappa,p)-\frac{3a_{2}^{2}(a_{2}^{2}-a_{1}^{2})}{\sqrt{a_{3}^{2}-a_{1}^{2}}} \;\Pi\left(\kappa,\frac{p^{2}a_{1}^{2}}{a_{2}^{2}},p\right)\right\},
\end{split}
\end{equation}

\noindent where $\kappa=\arcsin \sqrt{\frac{(a_{3}^{2}-a_{1}^{2})(a^{2}-a_{2}^{2})}{(a_{3}^{2}-a_{2}^{2})(a^{2}-a_{1}^{2})}}$ and $p=\sqrt{\frac{a_{3}^{2}-a_{2}^{2}}{a_{3}^{2}-a_{1}^{2}}}$. Substituting Eqs. (\ref{eq:PositiveLambdaPotentialFactorized}) and (\ref{eq:WKBforbiddenargument}) into Eq. (\ref{eq:forbiddenWKBwavefunction}), one finds the WKB wave function for $a_{1}<a_{2}<a<a_{3}$.

\subsection{WdW solution for $\Lambda<0$}

One is interested in the WKB wave function for the classically allowed region $a_{1}<a<a_{2}$. The WKB wave function is given by Eq. (\ref{eq:allowedWKBwavefunction}). The steps are the very ones  of the above procedure, however, following the discussion of Sec. \ref{sec:WdWEquation}, one must consider that the smaller root ($a^{2}$) is negative and real (cf. Eq. (\ref{eq:NegativeLambdaPotentialFactorized})). Using the formulas $3.131.5$, $3.132.4$, $3.137.5$ of Ref. \cite{Gradshteyn:1994} and the reduction formula $230.01$ of Ref. \cite{Byrd:1954}, one gets

\begin{equation}\label{eq:NegativeWKBforbiddenargument}\begin{split}
\int_{a_{1}}^{a} \sqrt{|V(a)|} da =& \frac{\sqrt{-g_{\Lambda}}}{3}\left\{\sqrt{(a_{2}^{2}-a^{2})(a^{2}-a_{1}^{2})(a^{2}+a_{i}^{2})} -\frac{(a_{1}^{2}+a_{2}^{2}-a_{i}^{2})(a_{1}^{2}+a_{i}^{2})}{\sqrt{a_{2}^{2}-a_{i}^{2}}}\;\Pi\left(\kappa,p^{2},p\right)+\right. \\
 & \left. \left(\frac{a_{i}^{4}+a_{i}^{2}a_{2}^{2}+a_{i}^{2}a_{2}^{2}-5a_{2}^{2}a_{1}^{2}}{\sqrt{a_{2}^{2}+a_{i}^{2}}}\right)F(\kappa,p)-\frac{3a_{2}^{2}(a_{1}^{2}+a_{i}^{2})}{\sqrt{a_{2}^{2}+a_{i}^{2}}} \;\Pi\left(\kappa,\frac{-p^{2}a_{i}^{2}}{a_{1}^{2}},p\right)\right\},
\end{split}
\end{equation}

\noindent where $\kappa=\arcsin \sqrt{\frac{(a_{3}^{2}-a_{1}^{2})(a^{2}-a_{2}^{2})}{(a_{3}^{2}-a_{2}^{2})(a^{2}-a_{1}^{2})}}$ and $p=\sqrt{\frac{a_{3}^{2}-a_{2}^{2}}{a_{3}^{2}-a_{1}^{2}}}$. The WKB wave equation is obtained after inserting Eqs. (\ref{eq:NegativeLambdaPotentialFactorized}) and (\ref{eq:NegativeWKBforbiddenargument}) into Eq. (\ref{eq:allowedWKBwavefunction}). This completes the WdW solutions for the potentials given by Eqs. (\ref{eq:NullLambdaPotential}) and (\ref{eq:NonullLambdaPotential}). Any analysis of the WKB wave function in this regime, due to its complex expressions is somewhat difficult. 

\section{Intepretation of the wave function}\label{sec:interpretation}

To analyse the behaviour of the wave function, one must compute $\mathcal{K}^{2}=K_{ij}K^{ij}$, the trace of the square of the extrinsic curvature Eq. (\ref{eq:extrinsiccurvature}) \cite{Bertolami:1998hm,Bertolami:1996pc}. If a wave function is oscillatory (exponential), $\mathcal{K}^{2}$ has positive (negative) eigenvalues. Using Eqs. (\ref{eq:Kij}) and (\ref{eq:canonicalmomentum}),  one gets

\begin{equation}
\mathcal{K}^{2}=-\frac{9}{\sigma^{2}a^{4}}\frac{d^{2}}{da^{2}}.
\end{equation}

Defining the auxiliary quantity $W:=\frac{\mathcal{K}^{2}\psi(a)}{\psi(a)}$ and the asymptotic expression Eq. (\ref{eq:HLdominantsolution}), for $a\ll1$, one obtains that $W<0$. When the HL gravity terms dominate the universe, the wave function is exponential, corresponding to an Euclidean geometry. 

For the very late universe, the behaviour of the wave function is very different depending on the value of the cosmological constant. If $\Lambda=0$, the wave function is given by Eq. (\ref{eq:curvaturedominantsolution}) and it is easy to show that $W<0$. Showing that the behaviour is exponential. If $\Lambda>0$, the wave function is given by a combination of oscillatory functions Eq. (\ref{eq:psipositivelambdafinal}), giving $W>0$, meaning that the geometry is Lorentzian or classical. Finally, for negative values of the cosmological constant, Eq. (\ref{eq:psinegativelambdafinal}) yields $W<0$. In the case studied here, the computations are quite simple, and it is not surprising to find this result since the nature of the wave function given by Eqs. (\ref{eq:HLdominantsolution}), (\ref{eq:curvaturedominantsolution}), (\ref{eq:psipositivelambdafinal}) and (\ref{eq:psinegativelambdafinal}) can be obtained directly by inspection. 

The semiclassical approximation implies that the configurations will oscillate about the classical solution \cite{Kiefer:2007}. In order to verify whether GR can be recovered for the low-energy limit in this approximation, one obtains the Hamilton-Jacobi equation from the WdW equation through the WKB method. The WdW equation (cf. Eq. (\ref{eq:wdwequation})) can be written as

\begin{equation}
 \left\{\frac{d^{2}}{d a^{2}}-V(a)\right\}\Psi(a)=0;
\end{equation}

\noindent substituting a wave function of the form $\psi=\mbox{Re} \left[Ce^{iS}\right]$,  where $C$ is a slowly varying amplitude and $S$ is the phase, one obtains the Hamilton-Jacobi equation

\begin{equation}\label{eq:HamiltonJacobi}
 \left(\frac{dS}{da}\right)^{2}+V(a)=0.
\end{equation}

\noindent Notice that $S$ is real provided $V(a)\leq0$ (classically allowed region), denoting that the wave function is oscillatory. If $V(a)>0$ (classically forbidden region), $S$ is imaginary and the wave function has an exponential behaviour. Using Eq. (\ref{eq:HamiltonJacobi}), one finds that $S=\int\sqrt{-V(a)}da$ is the phase for the classically allowed region. This integral was discussed in Sec. \ref{sec:WdWEquationSolutions}. Applying $\Pi_{a}$ to the wave function of the universe, $\psi=\mbox{Re} \left[Ce^{iS}\right]$, one gets $\Pi_{a}\psi=-i\frac{d\psi}{da}=-i\left(\frac{dC}{da}+iC\frac{dS}{da}\right)e^{iS}$. The WKB assumption, $\left|\frac{dC}{da}\right| \ll \left|\frac{dS}{da}\right|$, yields that $\Pi_{a}=\frac{dS}{da}$. Using Eqs. (\ref{eq:canonicalmomentum}) and (\ref{eq:HamiltonJacobi}), one is lead to

\begin{equation}\label{eq:scalefactort}
 t=\int_{a(0)}^{a(t)}da \frac{a}{\sqrt{-V(a)}},
\end{equation}

\noindent this equation relates the global time and the scale factor. One intends to investigate the regions where the positive cosmological constant dominates, so $V(a)\sim -g_{\Lambda} a^{4}$,  substituting this asymptotic potential into Eq. (\ref{eq:scalefactort}), one gets that the time evolution for the scale factor when $a\gg1$ is given by

\begin{equation}\label{eq:desitterevolution}
 a(t)\sim e^{\sqrt{g_{\Lambda}} t}.
\end{equation}

\noindent This corresponds to a de Sitter expansion phase, a behaviour expected for the GR regime, which is recovered when $\lambda=1$. Of course, this does not prove that GR is recovered as an IR fixed point of the HL gravity, but shows that a HL FLRW cosmology yields for $a\gg1$ a semiclassical solution that corresponds to the GR one. A possible approach to tackle the problem would involve considering a scaling $\lambda=\lambda(a)$, and expect that $\lambda\rightarrow 1$ for $a\gg 1$. Considerations of this nature were developed for cosmological models with scale-dependent Newtonian gravitational coupling (cf. Refs. \cite{Bertolami:1993mh,Bertolami:1995rt}).

\section{Conclusions}\label{sec:conclusions}

In this work, the quantum cosmology for the Ho\v{r}ava-Lifshitz gravity without matter is investigated for the closed universe. In the minisuperspace model, the WdW equation is derived, and it is shown that the HL gravity introduces terms that are dominant for short distances, modifying the behaviour of GR on these scales. One chooses the configurations for which the HL gravity new terms act as a ``potential barrier''  close to the singularity, $a=0$.

The solutions for the WdW equation are obtained considering the deWitt boundary condition, Eq. (\ref{eq:dWBC}), which states that the wave function vanishes at the singularity. For $a\ll1$, corresponding to the very early universe when the HL gravity terms dominate, the wave function is an exponential,  that is typical of  classically forbidden region. A quantum bound for the coefficient $g_{s}$ is found. 

For the very late universe, when $a\gg1$, the curvature and the cosmological constant terms dominate and, one finds that, for $\Lambda=0$ or $\Lambda<0$, the wave function is exponentially suppresed, denoting as before, that this region is classically forbidden. For a positive cosmological constant case, one finds a damped oscillatory behaviour, already found in the usual QC for GR. 

For the vanishing cosmological constant, an exact solution can be obtained. In this case, one finds that the singularity is avoided due to quantum effects as the probability to reach the singularity  $a=0$ vanishes and that $g_{r}$ is quantized. Fixing the value of $g_{s}$, for large values of $n$ (large $g_{r}$ values), one can obtain a classical universe according to the analog of the correspondence principle of old quantum mechanics. The complete exact solution for $\Lambda\neq 0$ cannot be obtained, although wave functions in the WKB approximation can be obtained for the intermediate regions.

Finally, the discussion on how the classical solution emerges from the semiclassical analysis is performed solving the Hamilton-Jacobi equation: one encounters a semiclassical solution oscillating nearby the classical solution. For $\Lambda>0$ and $a\gg1$, this leads to a de Sitter space-time, as expected from GR.

One then concludes that quantum cosmology applied to HL gravity suggests that this proposal matches the expectations of a quantum gravity model for the very early universe, as it provides, for instance, a hint for the singularity problem for the $\Lambda=0$ case. In what concerns specific solutions, the model suggests that GR behaviour is recovered at the semiclassical limit.


\section*{Acknowledgments}


C. Z. would like to thank Funda\c{c}\~{a}o para a Ci\^{e}ncia e a Tecnologia
(FCT)  for the financial support under the fellowship SFRH/BPD/75046/2010 and the FCT project CERN/FP/116358/2010; Centro de F\'{i}sica do Porto for the hospitally during the period that this paper was prepared and Miguel Costa for stimulating discussions about some issues related with this work. The work of O. B. is partially supported by the FCT projects PTDC/FIS/111362/2009 and CERN/FP/116358/2010.




\end{document}